\documentclass[fleqn,usenatbib]{mnras}

\usepackage{newtxtext,newtxmath}
\usepackage[T1]{fontenc}

\DeclareRobustCommand{\VAN}[3]{#2}
\let\VANthebibliography\thebibliography
\def\thebibliography{\DeclareRobustCommand{\VAN}[3]{##3}\VANthebibliography}

\usepackage{graphicx}	% Including figure files
\usepackage{amsmath}	% Advanced maths commands
\newcommand{\tess}{{\it TESS}}
\newcommand{\asassn}{{\it ASAS-SN}}

\newcommand{\gaia}{{\it Gaia}}
\newcommand{\xmm}{{\it XMM-Newton}}
\newcommand{\chandra}{{\it Chandra}}
\newcommand{\mass}{{\it 2MASS}}
\newcommand{\wise}{{\it WISE}}

\title[The Peculiar Bursting Nature of CP Pup]{The Peculiar Bursting Nature of CP Pup}

\author[M. Veresvarska et al.]{
M. Veresvarska,$^{1}$\thanks{E-mail: martina.veresvarska@durham.ac.uk}
S. Scaringi,$^{1}$
S. Hagen,$^{1}$
D. De Martino,$^{2}$
C. Done,$^{1}$
K. Ilkiewicz, $^{1,3}$
\newauthor
\ C. Knigge, $^{4}$
C. Littlefield$^{5}$
\\
$^{1}$Centre for Extragalactic Astronomy, Department of Physics, Durham University, South Road, Durham, DH1 3LE\\
$^{2}$INAF-Osservatorio Astronomico di Capodimonte, Salita Moiariello 16, I-80131 Naples, Italy\\
$^{3}$Astronomical Observatory, University of Warsaw, Al. Ujazdowskie 4, 00-478 Warszawa, Poland\\
$^{4}$School of Physics and Astronomy, University of Southampton, Highfield, Southampton SO17 1BJ, UK\\
$^{5}$Bay Area Environmental Research Institute, Moffett Field, CA 94035, USA\\
}

\date{Accepted XXX. Received YYY; in original form ZZZ}

\pubyear{2023}

\begin{document}
\label{firstpage}
\pagerange{\pageref{firstpage}--\pageref{lastpage}}
\maketitle

\begin{abstract}
The classical nova CP Puppis has been observed to have particularly puzzling and peculiar properties. In particular, this classical nova displays occasional bursts in its long-term \asassn\ light curve. Here we report on 5 sectors of \tess\ data displaying 2 of these rapid bursts, lasting $\sim$ 1 day. Based on the estimated lower energy limits of the bursts we discuss whether the bursts may be examples of micronovae resulting from localised thermonuclear explosion. Furthermore, its orbital period remains uncertain, with several inconsistent periodic signals appearing in spectroscopic and photometric observations at various wavelengths. Although we cannot unambiguously unravel the physical origin of the signals, the previously suggested nature of CP Puppis as a long orbital period system may be a viable explanation. The recurrence time of the bursts in CP Puppis, together with the unexplained variable modulations make it a prime candidate for intense monitoring.
\end{abstract}


%%%%%%%%%%%%%%%%%%%%%%%%%%%%%%%%%%%%%%%%%%%%%%%%%%

%%%%%%%%%%%%%%%%% BODY OF PAPER %%%%%%%%%%%%%%%%%%

\section{Introduction}
\label{s:intro}
CP Puppis (CP Pup hereafter) is a well studied cataclysmic variable that has undergone a nova explosion in 1942. It is an unusually bright and fast example of a nova explosion, with the difference in amplitude of apparent magnitude of $\sim 17$ mag \citep{Payne-Gaposchkin1964}. It is also a particularly fast nova with $t_{3} \sim 6.5$ d \citep{Payne-Gaposchkin1964}, where $t_{3}$ represents the time it takes for the nova luminosity to decline by 3 magnitudes. It has been reported that since the nova explosion the brightness levels of the nova have yet to return to the pre-burst quiescence level \citep{Schaefer2010}. Similarities can be drawn between CP Pup and V1500 Cyg \citep{DellaValle1998}, which is also uncharacteristically brighter and has also remained brighter post eruption. CP Pup is also suggested to have a magnetic accretor \citep{Balman1995,Orio2009,Mason2013}. 

One of the main peculiarities of CP Pup is its elusive orbital period. There have been spectroscopic \citep{Bianchini1985,Duerbeck1987,ODonoghue1989,White1993,Bianchini2012,Mason2013} and photometric \citep{Warner1985,ODonoghue1989,Diaz1991,White1993,Bruch2022} periods reported between $0.06115$ days and $0.06834$ days. A summary of all these is provided in \citet{Bruch2022}. Both spectroscopic and photometric periods reported in literature lie in a similar range, with most of the spectroscopic periods being shorter. Furthermore, the spectroscopic periods have been reported to show large scatter in the folded radial velocity curve \citep{Bianchini2012}. It is worth noting that the spectroscopic and photometric periods in the literature are inconsistent with each other \citep{Diaz1991}. \citet{Orio2009} has further reported 3 periods from \xmm\ data. However, due to the large uncertainty, the periods quoted in \citet{Orio2009} are consistent with each other, as well as consistent with previously reported spectroscopic and photometric periods (see summary by \citet{Bianchini2012} and \citet{Bruch2022}). 
%The periods determined in \citet{Orio2009} are all reported to be consistent with each other to $1 \sigma$ and encompass most of the previously reported periods as well. A detailed history of the spectroscopic and photometric period measurements is provided in \cite{Bianchini2012} and \cite{Bruch2022}. 

If taken at face value as the orbital period, all the reported values are short for known classical novae and non-magnetic nova-likes in general ($\sim 3$ hours), making CP Pup one of a handful of novae with a potential period reported below the period gap \citep{Bianchini1985,Marelli2018}. It has thus been suggested that CP Pup is a system whose disc is always hot and bright, and often exhibits superhumps, persisting in a state similar to SU UMa type dwarf novae during superoutburst \citep{Warner1985,Patterson1998,Patterson2013}.

Further interpretations of the period assigned it to the spin of the magnetic white dwarf, suggesting a white dwarf slightly out of spin-orbit synchronism \citep{Warner1985,Diaz1991,Balman1995}. Asynchronous polars are not the only option for the magnetic interpretation of CP Pup. \citet{Orio2009} and \citet{Mason2013} consider the hypothesis of CP Pup being an intermediate polar. However, there is a lack of conclusive evidence for either interpretation.

The peculiar nature of CP Pup extends to other observations too. The relatively recent nova explosion, with its characteristics such as peak absolute magnitude, suggests a high mass white dwarf (WD), due to the smaller physical accretor size resulting in higher pressure, making it more favourable to trigger a nova explosion \citep{Prialnik1995}. However, \citet{White1993} points out that the reported values of $K_{1} \sim 70 - 146$ kms$^{-1}$ require a WD mass $< 0.2 M_{\odot}$. This is in contradiction to the nova nature of CP Pup. Some more recent studies adopting higher WD mass of 1.1 $M_{\odot}$ have instead obtained better fits to optical spectra with an accretion rate in the range of $3.3 - 7.3 \times 10^{-10} M_{\odot}yr^{-1}$ \citep{Mason2013}. Also assuming a high mass WD, $M_{WD} > 1.1 M_{\odot}$, \citet{Orio2009} derived an upper limit on mass transfer rate $\lesssim 1.6 \times 10^{-10} M_{\odot}yr^{-1}$. However, this estimate assumes a distance of 1600 pc, more than double the distance of 780 $\pm$ 11 pc inferred from the \gaia\ parallax \citep{Gaia2016,Gaia2023}. \citet{Orio2009} also quotes a $3\sigma$ upper limit $\lesssim 8 \times 10^{-11} M_{\odot}yr^{-1}$ at 850 pc, as determined by \citet{Cohen1983}. Other estimates of the accretion rate adopting the more precise \gaia\ parallax measurement have been made. \citet{Selvelli2019} reports a considerably higher value of $10^{-9.22\pm0.32}$ $M_{\odot}$yr$^{-1}$. This is related to the unusually high extinction of $A_{v} \sim 0.6$ mag, which they adopt.

In this work we report on the analysis of 5 sectors of \tess\ data of CP Pup (Section \ref{s:obs}). In Section \ref{s:res} we discuss 2 rapid bursts observed as well as  the phenomenology of the cluster of periods found in the \tess\ light curves. Reanalysis of \xmm\ data of CP Pup is also discussed here and the new WD mass estimate is derived. In Section \ref{s:conclusion} we discuss the observed rapid bursts in the context of the micronova model (\citet{Scaringi2022a,Scaringi2022}.
\section{Observations}
\label{s:obs}

The data used in this work was obtained by \tess\ and is available on the Mikulski Archive for Space Telescopes (MAST\footnote[1]{\label{mast}\url{https://mast.stsci.edu/portal/Mashup/Clients/Mast/Portal.html}}). \tess\ observed CP Pup over 5 sectors amounting to $\sim$ 5 months of observations at 2 minute cadence. The last sector 61 also contains 20 second cadence data. The detailed description of the data with the burst characteristics is provided in Table \ref{tab:obs}. \tess\ provides 2 modes for output flux. Here the SAP flux is used, as opposed to the processed version of PDCSAP. This is to avoid any contamination of the data through pipeline processing. The flux is then cleaned with a mask discarding any data points with quality flag $>$ 0, to mitigate presence of artificial or non-intrinsic features in the data.

\begin{figure*}
	\includegraphics[width=\textwidth]{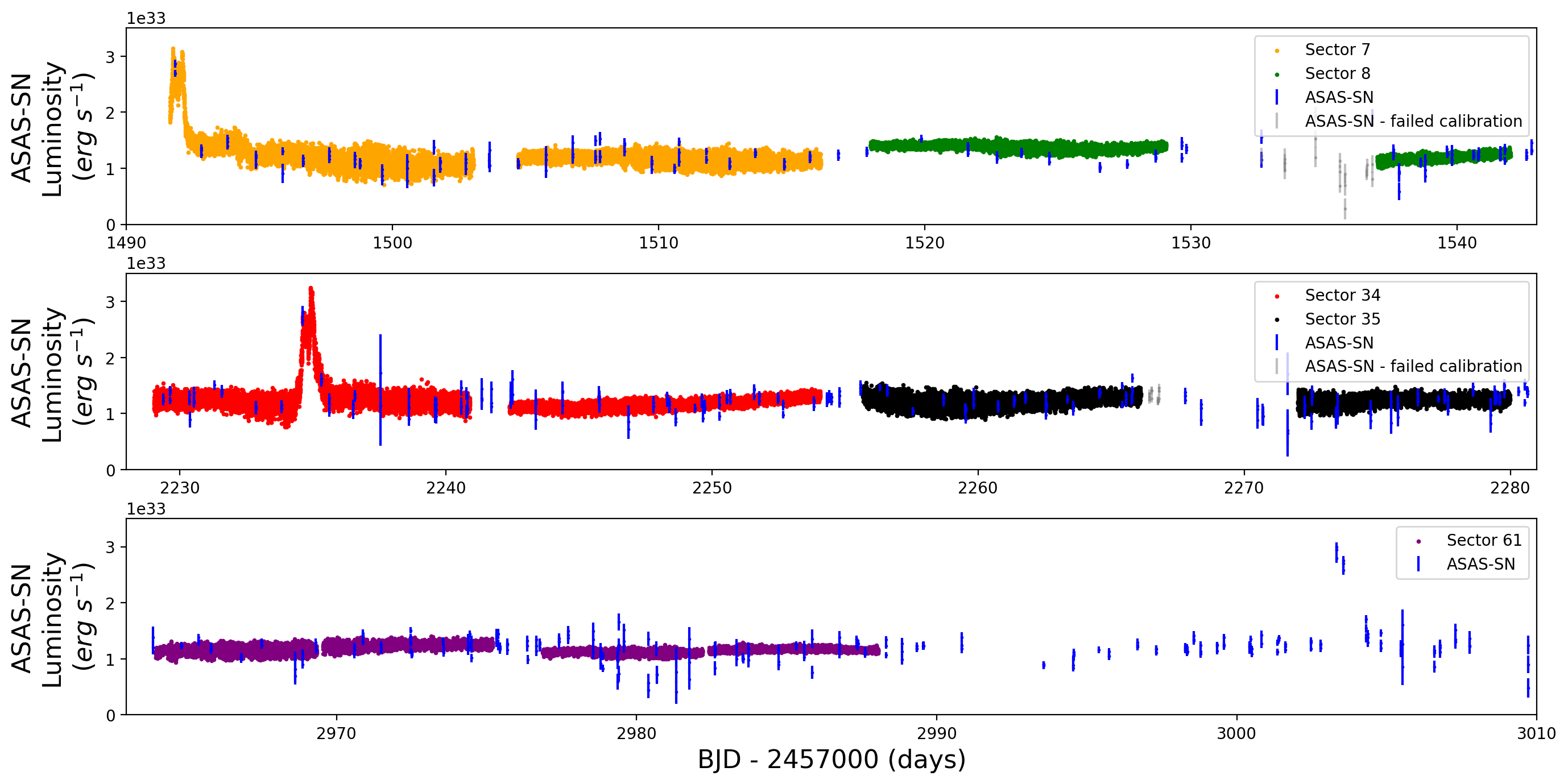}
    \caption{Light curve of CP Pup showing all 5 sectors of \tess\ data as described in Table \ref{tab:obs} (orange, green, red, black and purple consecutively for sectors 7, 8, 34, 35 and 61). Overlaid (in blue) is the \asassn\ light curve used for calibration. The grey points are data from \asassn\ for which calibration failed and hence were excluded. The flux of both \tess\ and \asassn\ light curves is converted to erg s$^{-1}$ using the \gaia\ parallax.}
    \label{fig:LC}
\end{figure*}

The \tess\ data shows a modulation of the shape of the light curve in the form of a larger envelope, where the variance and flickering temporarily varies on timescales of $\lesssim 10$ days (see Fig. \ref{fig:LC}). The same envelope is present to a greater extent in the 20 s cadence data of Sector 61 in \tess\ . The Lomb-Scargle periodogram of the 20 s data shows a QPO-like feature at $\approx1500$ cycles$/$day. This is also present in all the neighbouring stars in the field of CP Pup and hence assumed to not be intrinsic to the CP Pup itself, but rather a common systematic. Similar behaviour is also observed in other \tess\ 20 s observations in Sector 61 (such as TIC 260266504, TIC 289113766 and TIC 289113764). This points to an instrumental origin of the QPO and is hence disregarded. As no other new signals are found in the 20 s cadence data of Sector 61, the 2 minute cadence is used throughout for consistency. The data is downloaded and cleaned for cosmic rays through the \texttt{Lightkurve} package\footnote[2]{\url{https://docs.lightkurve.org/index.html}}. 

\begin{table}
	\centering
	\caption{Summary of the \tess\ data of CP Pup, with sector numbers and dates. If a sector contains a burst, the corresponding date in Barycentric Kepler Julian date, duration in days and energy is noted. The details of energy calculation are described in Section \ref{ss:bursts}.}
	\label{tab:obs}
	\begin{tabular}{lcccr} % four columns, alignment for each
		\hline
		Sector & Dates & burst dates & Duration & Energy \\
         & (BTJD-2457000) & (BTJD-2457000) &  (d) & (erg) \\
		\hline
		7 & 1491.6 - 1516.1 & 1491.6 - 1492.2 & $\sim 0.6$ & $1.2\times 10^{38}$\\
		8 & 1518.0 -  1542.0 & $-$ & $-$ & $-$\\
		34 & 2229.0 -  2254.1 & 2229.0 - 2234.4 & $\sim 1$ & $6.3\times 10^{37}$\\
        35 & 2255.7 -  2280.0 & $-$ & $-$ & $-$\\
        61 & 2964.0 - 2988.1 & $-$ & $-$ & $-$\\
		\hline
	\end{tabular}
\end{table}

While \tess\ provides an excellent precision relative photometry, it is better to leverage data from another observatory to obtain absolute photometry from \tess\ data. To convert the \tess\ data to erg s$^{-1}$ it is necessary to consider ground based observations as well as knowledge of its distance. Quasi-simultaneous \asassn\ g-band data \citep{Shappee2014,Kochanek_2017} is used for this purpose. The band is centered at $475$ nm with $140$ nm width, with a partial overlap with the \tess\ band-pass ($600$ - $1000$ nm). Although the pass-bands do not exactly overlap, we here assume that any colour-term variation are minimal. The data is obtained from the \asassn\ webpage\footnote[3]{\label{asassn}\url{https://asas-sn.osu.edu/}}. The conversion is done through simultaneous observations for each half sector. Simultaneous data consist of all \asassn\ observations within a \tess\ cadence observation of 2 minutes. Assuming a linear relation between the 2 bands a direct conversion can be established in the form of $F_{ASAS-SN } \left[ mJy \right] = A \times F_{TESS} \left[ e^{-}s^{-1} \right]  + C$. The corresponding values of the coefficients for each half-sector are specified in Table \ref{tab:cal} to account for any deviations in data due to the gap in the middle of \tess\ sectors.

\begin{table}
	\centering
	\caption{Summary of the conversion coefficients from \tess\ flux in $e^{-s}$ to \asassn\ flux in mJy for CP Pup. As the conversion is done twice per \tess\ sector, all corresponding coefficients are listed.}
	\label{tab:cal}
	\begin{tabular}{lccr} % four columns, alignment for each
		\hline
		Sector & Sector half & A $\bigg( \frac{mJy}{e^{-}s^{-1}} \bigg)$ & C (mJy) \\
		\hline
		7 & 1 & 0.0241 $\pm$ 0.0007 & -1.2 $\pm$ 0.1\\
          & 2 & 0.016 $\pm$ 0.002 & -0.0 $\pm$ 0.3\\
		8 & 1 & 0.011 $\pm$ 0.001 & -0.9 $\pm$ 0.4\\
		   & 2 & 0.011 $\pm$ 0.001 & -0.9 $\pm$ 0.4\\
        34 & 1 & 0.0209 $\pm$ 0.0008 & -11.1 $\pm$ 0.5\\
           & 2 & 0.011 $\pm$ 0.001 & -4.6 $\pm$ 0.9\\
        35 & 1 & 0.015 $\pm$ 0.001 & -2.2 $\pm$ 0.4\\
           & 2 & 0.014 $\pm$ 0.003 & -1.8 $\pm$ 0.9\\
        61 & 1 & 0.013 $\pm$ 0.001 & -1.8 $\pm$ 0.4\\
           & 2 & 0.008 $\pm$ 0.002 & -0.5 $\pm$ 0.7\\
		\hline
	\end{tabular}
\end{table}

There are 2 cases where the calibration between \tess\ and \asassn\ fails. The first is in Sector 8, where the \tess\ light curve shows a drop in flux inconsistent with the ground-based \asassn\ observations. This is also reported in \citet{Bruch2022}. Another similar case occurs in Sector 35 where the \tess\ data shows an uncharacteristic rise in flux, but this appears to be shared within the entire region surrounding CP Pup and we attribute this to poor background correction. Consequently both segments of Sectors 8 and 35 where the calibration to \asassn\ fails are discarded. The final result is shown in Figure \ref{fig:LC}, where the light curve is also corrected for the distance of $780 \pm 11$ pc inferred from \gaia\ DR3 parallax. Despite the high extinction in \gaia\ passband, no bolometric correction has been applied. This is to avoid any effects due to a potential change in the SED between quiescent and bursting parts of the light-curve, which would affect any bolometric correction. As a result the obtained luminosity of CP Pup should be treated as a lower limit. It is however possible to estimate the effect the bolometric correction could have on the luminosity. The X-ray luminosity in Section \ref{ss:SEDx} is estimated to be $L_{X} ~\sim 1.1 \times 10^{33}$ erg s$^{-1}$. Similarly, \citet{Orio2009} estimated the UV luminosity to be $L_{UV} \sim 2 \times 10^{33}$ erg s$^{-1}$. Alongside with the mean \tess\ luminosity of $L_{ \tess\ } \sim 1 \times 10^{33}$ erg s$^{-1}$ the total bolometric luminosity is expected to be of the order of $\sim 4 \times 10^{33}$ erg s$^{-1}$. Therefore the lower limits on the luminosity are expected to represent about $\sim 4 \times$ lower values than the total bolometric values.

Further data used in this work has been obtained by \xmm\ on 4$^{th}$ of June 2005 and reported in \citet{Orio2009}. As described in greater detail in \citet{Orio2009}, the data was obtained with EPIC-pn, MOS-1 and MOS-2 with $\sim 50$ ks exposure. Grating spectrum with low SNR was also obtained with RGS and reported in \citet{Orio2009}. The spectrum is shown in \citet{Orio2009} in Figure 3. The reduced data is available online at {\tt heasarc}\footnote[4]{\url{https://heasarc.gsfc.nasa.gov/cgi-bin/W3Browse/w3browse.pl}}. However, the data reanalysed here, is only the data taken with the most sensitive EPIC-pn instrument.

\section{Results}
\label{s:res}

In this section we report the bursts found in 2 of the 5 available \tess\ sectors in Section \ref{ss:bursts}. We further report on the observed variable cluster of frequencies found in the power spectrum of CP Pup in Section \ref{ss:cluster}. The \xmm\ spectrum reported in \citet{Orio2009} and our new model fit is presented in Section \ref{ss:SEDx}

\subsection{Bursts}
\label{ss:bursts}
\begin{figure*}
	\includegraphics[width=\textwidth]{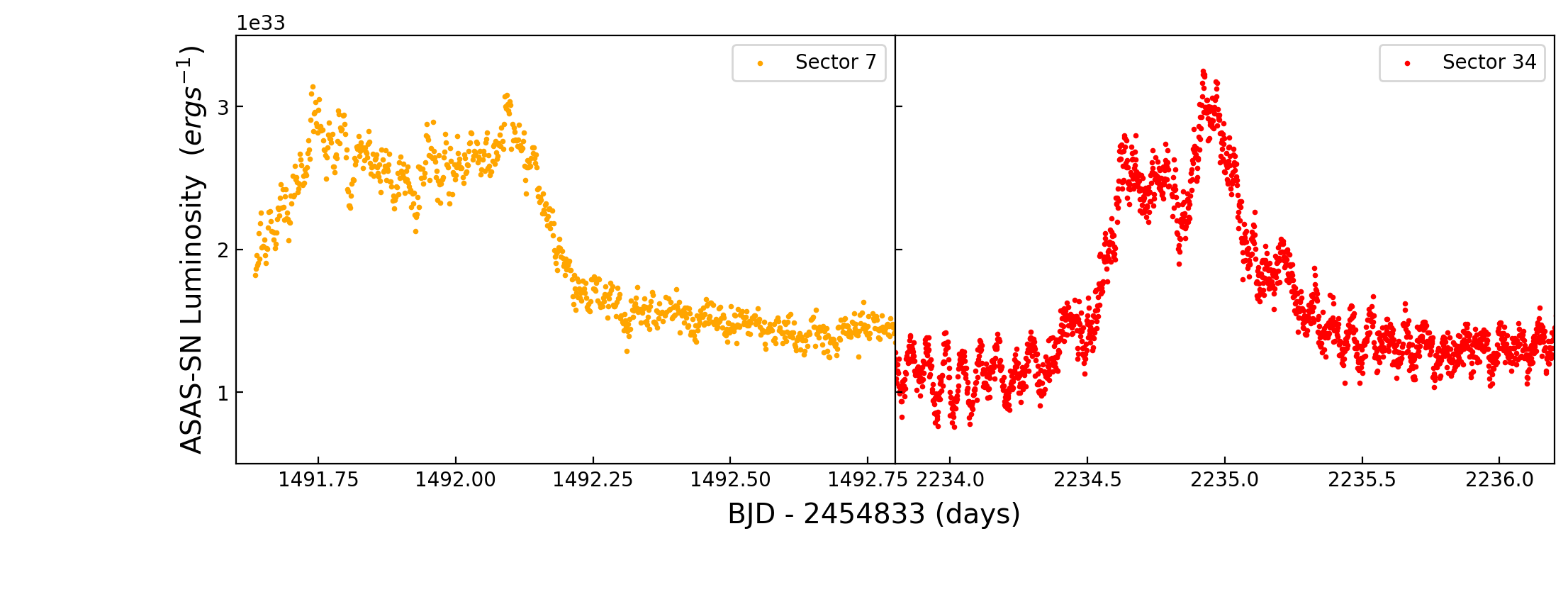}
    \caption{Zoom in on the bursts shown in Figure \ref{fig:LC} and described in Table \ref{tab:obs}. The energy of each burst is $1.2 \times 10^{38}$ erg for sector 7 and $6.3 \times 10^{37}$ erg for sector 34.}
    \label{fig:bursts}
\end{figure*}

The \tess\ light curve from all sectors is displayed in Figure \ref{fig:LC}, where two bursts are detected in Sector 7 and 34 with the corresponding times in Table \ref{tab:obs}. The bursts have also been reported in \citet{Bruch2022}, however they are disregarded in the analysis. The recorded bursts in Sectors 7 and 34 all lasted $\lesssim 1 $ day, with the overall amplitude increasing by a factor of $>2$, from $\sim 1.5 \times 10^{33}$ erg s$^{-1}$ to $> 3.0 \times 10^{33}$ erg s$^{-1}$. A detailed version of the bursts is shown in Figure \ref{fig:bursts}. Only Sector 34 contains both the immediate pre-burst and post-burst observation. Sector 7 started during the rise of the burst and contains the post-burst part only. It is worth noting an additional burst of similar nature in the long term \asassn\ light curve at $\sim 3005$ BTJD in Figure \ref{fig:LC}. Unfortunately there are no \tess\ observations around that date. Another interesting characteristic of the bursts is their shape, showing 2 distinct peaks with $\sim$ 0.35 d separation.

To understand the origin and nature of the bursts, their energy has to be determined. This is done by integrating the luminosity under the calibrated light curve after subtracting the baseline luminosity level. We estimate the baseline luminosity using the pre- and post burst data only. We use these to compute a running mean and interpolate in between the two with a spline function. The resulting energy in ergs is reported in Table \ref{tab:obs}. The most energetic burst in Sector 7 releases in excess of $1.2 \times 10^{38}$ erg. The energy released by the burst in Sectors 34 is $>6.3 \times 10^{37}$ erg. We note that the aforementioned data gaps, as well as the lack of a bolometric correction allows us to only provide lower limits to the burst energies.

\subsection{Phenomenology of the period cluster}
\label{ss:cluster}

A peculiar feature of CP Pup is the cluster of periods between $\sim 14$ cycles$/$d and $\sim 15$ cycles$/$d ($\sim 1.5 - 1.6$ hr). In the 5 \tess\ sectors reported here all show this variability. This time-dependent variability is discussed in detail in \citet{Bruch2022}. The persistent signal reported from 4 \tess\ sectors in \citet{Bruch2022} is at $\sim$ 16.29 cycles$/$d. Independently redoing the period analysis on all 5 \tess\ sectors also reveals a single identical persistent signal in the cluster at 16.29 $\pm$ 0.04 cycles$/$d.

\begin{figure}
	\includegraphics[width=\columnwidth]{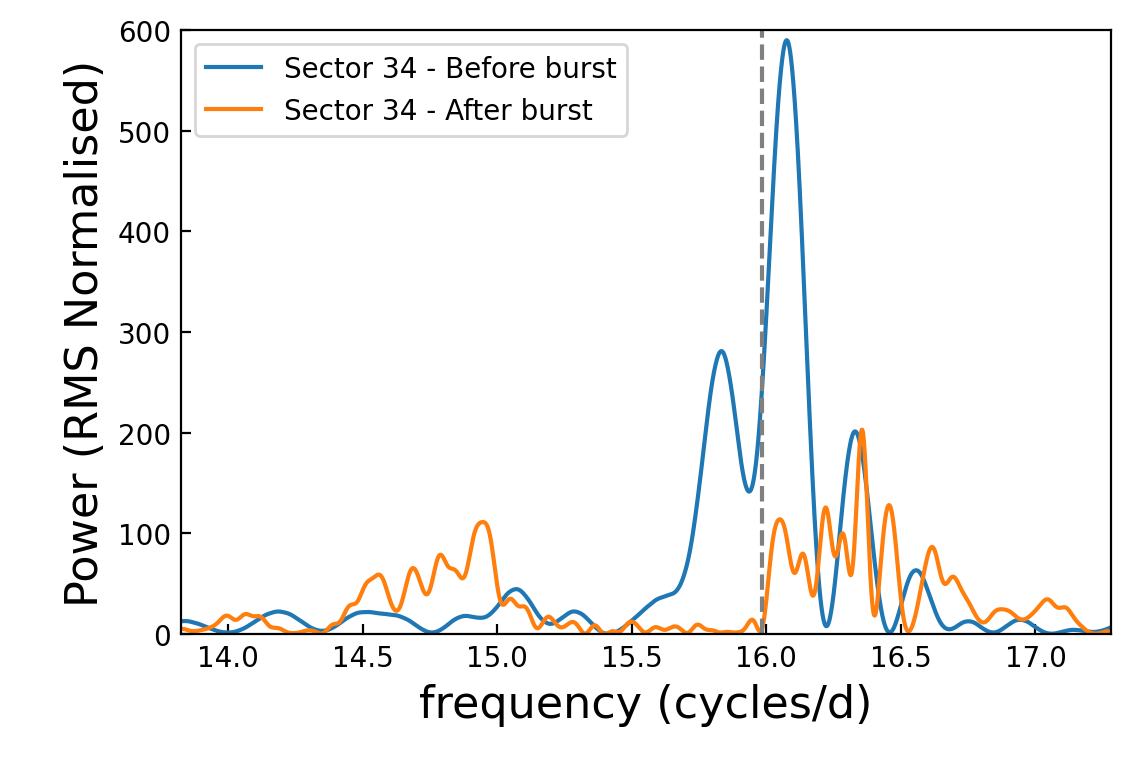}
    \caption{PSD of sectors 34 divided into pre-burst (blue) and post-burst (orange) parts. Grey dashed line shows the coherent \xmm\ period at $\sim 16$ cycles$/$d.}
    \label{fig:PSDburst}
\end{figure}

In Figure \ref{fig:PSDburst} the cluster is shown right before and after the burst present in sector 34. There is no apparent connection between the burst and the position or shift of the cluster frequencies. However, it is evident that there is considerably more rms power in the cluster before the bursts than after. For sector 34 the peak power of the cluster is at $\sim 590$ RMS, whereas after burst it is at $\sim 110$ RMS, excluding the power of the signal at $\sim 16.29$ cycles$/$d at $\sim 200$ RMS. Similarly, sector 7 shows a significant quasi-periodic variability as well. However lack of data pre-burst does not allow for similar comparison as in sector 35.

\citet{Orio2009} also reports only one coherent period in the light curve of CP Pup of 16 $\pm$ 2 cycles$/$d ($\sim$90 minutes). We recover the same signal from the \xmm\ data. The signal is marked by the dashed grey line in Figure \ref{fig:PSDburst}. Furthermore, this period also corresponds to the only persistent signal in \tess\ light curve.

%\begin{figure}
	%\includegraphics[width=\columnwidth]{CP_Pup_PSD_zoom_in.pdf}
    %\caption{PSD of each sector of \tess\ data of CP Pup, zoomed in on the frequencies between $14$ cycles$/$d and $17$ cycles$/$d. The zoomed in section showcases the period cluster and its variability through the available data. The persistent period is shown at $16.29$ cycles$/$d, with the variable components mostly centered on $\sim 15.5$ cycles$/$d. Sector 7 displays an aberrant power peak at $\sim 17$ cycles$/$d. Grey dashed line marks the frequency corresponding to $\sim 16$ cycles$/$d, the coherent period extracted from \xmm\ data from \citet{Orio2009}.}
    %\label{fig:PSDzoom}
%\end{figure}

\subsection{The potential nature of the orbital period}
\label{ss:Porb}

\begin{figure*}
	\includegraphics[width=\textwidth]{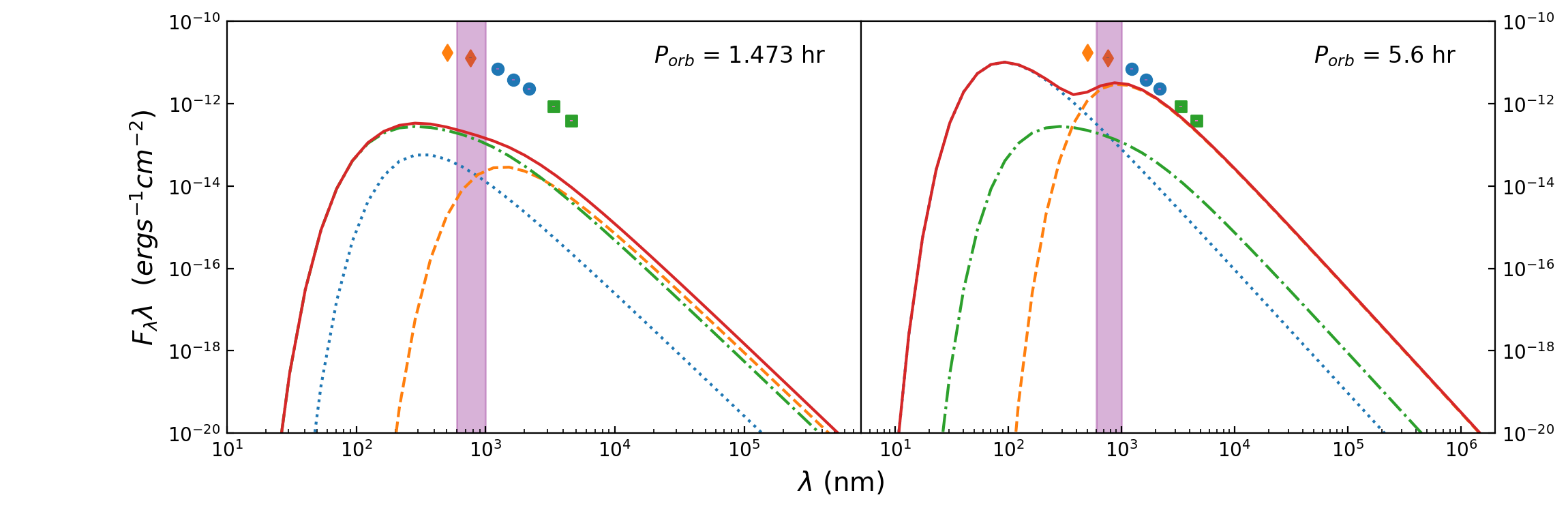}
    \caption{\textit{Left:} model SED of CP Pup for an estimated orbital period of 1.473 hr with the dotted (blue) line representing the WD, dashed (orange) line the secondary and the dashed-dotted (green) line the accretion disc. The total SED is shown in solid (red) line. For comparison the measurements from \wise\ W1 and W2 are shown (green squares) as well as the \mass\ filters (blue dots) and the \gaia\ measurements (orange diamonds). The \tess\ bandpass is shown in the shaded (pink) region. \textit{Right:} Same SED components and data as in the left panel but for orbital period of $\sim$ 5.6 hr, the longest orbital period included in the evolutionary tracks in \citet{Knigge2011}. }
    \label{fig:SED}
\end{figure*}

The signal at 1.473 hr is interpreted in literature as the orbital period of CP Pup. Using the evolutionary tracks of \citet{Knigge2011} we can estimate the temperature of the donor to be $\sim 2700$ K with radius of $\sim$ 0.131 $R_{\odot}$ for a system with an orbital period of $\sim 1.473$ hr. The corresponding white dwarf temperature according to \citet{Knigge2011} would be 11 000 K with radius following from the WD mass in Section \ref{ss:SEDx}. In such a case the SED of the system can be roughly approximated by two spheres radiating as blackbodies with the corresponding radii and temperatures. Another possible dominant SED component would be the accretion disc. Taking the X-ray derived accretion rate of $\sim$ 1 $\times$ 10$^{-10}$ $M_{\odot}$yr$^{-1}$ (Section \ref{ss:SEDx}) at face value, a standard Shakura-Sunyaev \citep{Shakura1973} accretion disc from the WD surface to the tidal radius at $\sim$0.31 $R_{\odot}$ is assumed. The resulting SED is hence shown in the left hand side panel of Figure \ref{fig:SED}. It is evident that there is a deficit in flux across all observed wavebands that cannot be corrected by either an increase in accretion rate or by correcting for extinction ($A_{V} \sim 0.6$ mag \citep{Selvelli2019}, $A_{H} = $0.107 mag and $A_{J} = $0.173 mag \citep{Cardelli1989} and $A_{W1} = $0.203 mag and $A_{W2} = $0.157 mag \citep{Fitzpatrick1999,Indebetouw2005}.

\citet{Mason2013} discusses the possibility that CP Pup may be a long orbital period CV. Despite no other photometric signal being detected, \citet{Mason2013} reports a substantial power at low frequencies ($\sim 2.38$ cycles/d). Their H$\beta$ RV curve produces a fit of 9.8 hr. They point out a large uncertainty of 1 day, making this a lower limit on the potential signal. The longest orbital period we can use from the evolutionary tracks of \citet{Knigge2011} at an orbital period of 5.6 hr provides an estimated donor temperature of $\sim$ 3900 K and radius of $\sim$ 0.63 $R_{\odot}$. Assuming the same parameters for the WD and accretion disc as previously (with the slight correction for tidal radius being at $\sim$ 0.19 $R_{\odot}$) the estimated SED is as shown in the right panel of Figure \ref{fig:SED}. This is more consistent to the observed \mass\ and \wise\ absolute magnitudes of CP Pup, suggesting a hotter and larger donor than previously anticipated. Since the evolutionary tracks in \citet{Knigge2011} do not include orbital periods above 5.6 hr, the right panel in Figure \ref{fig:SED} represents an estimate, not a fit, of the effect a longer orbital period would have on the SED. This also tentatively suggests a longer orbital period for CP Pup than reported in the past, but a more intensive radial velocity campaign is required to unravel the problem of the orbital period of CP Pup.

\subsection{Revisiting the WD mass and mass accretion rate with X-Rays}
\label{ss:SEDx} 
The reanalysis of the X-ray spectrum in this section is motivated by the inconsistent estimates of the WD maximal temperature $T_{\mathrm{max}}$ in the literature \citep{Orio2009,Mason2013}, leading to an uncertain WD mass and unconstrained accretion rate. This is also exacerbated by the pre-\gaia\ distances used in past X-ray analyses. More specifically \citet{Orio2009} infer large values of $T_{\mathrm{max}}$ following from the used models for non-magnetic systems and a simplified model of magnetic systems. Exploring the possibility of micronova eruptions present in CP Pup, the model implemented in this work is adapted specifically for magnetic accreting WD with an accretion column.

\begin{figure}
    \centering
    \includegraphics[width=\columnwidth]{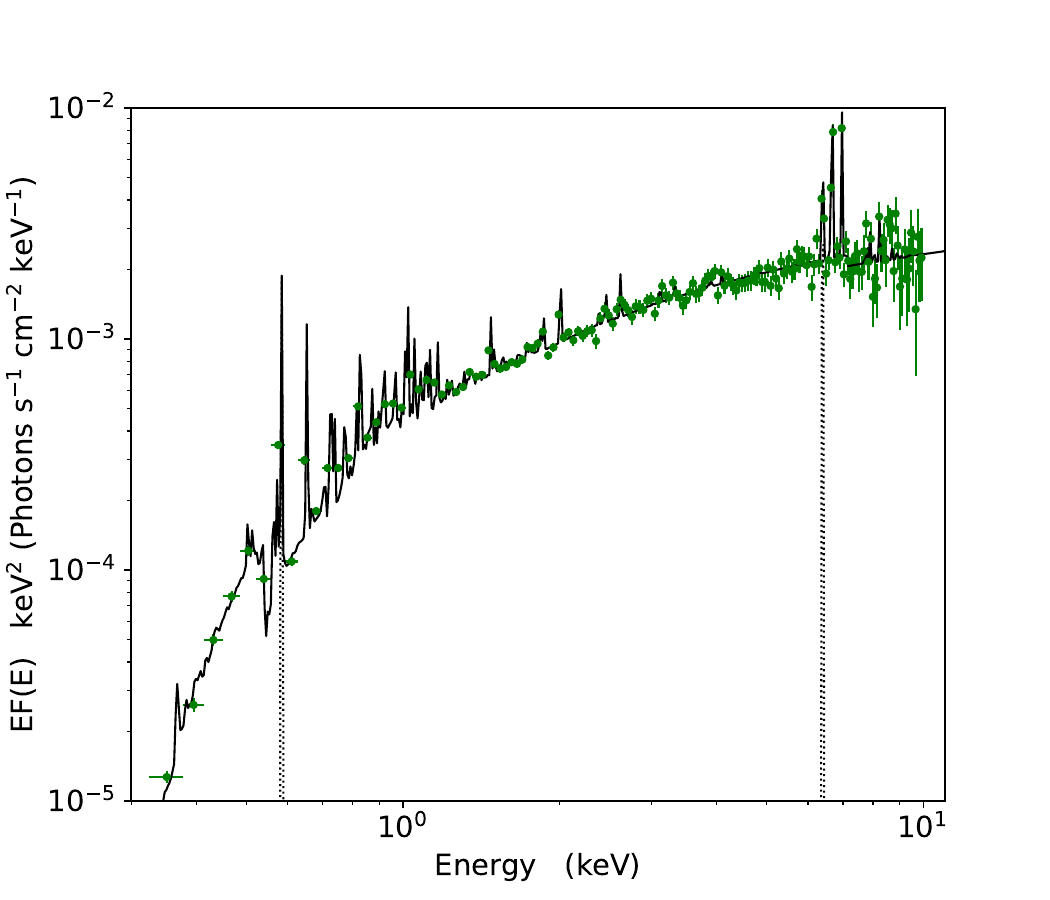}
    \caption{Unfolded XMM Epic-PN X-Ray spectrum of CP Pup. The green points show the data, while the solid black line shows the total model. The dotted gaussians show the two additional lines needed to fit the spectrum. These are (from left to right): O\,{\sc vii} and Fe-K$\alpha$}
    \label{fig:SEDx}
\end{figure}

Figure \ref{fig:SEDx} shows the archival \xmm \,spectrum of CP Pup taken on 4th June 2005, reported in \citet{Orio2009}. \citet{Orio2009} fit the RGS spectrum using {\sc mkcflow} model \citep{Mushotzky88} model which describes the emission from a cooling multi-temperature plasma and the EPIC-pn spectrum with a multi-temperature APEC model. However their analysis  does not include the combination with a complex absorber which is instead characteristic of magnetic CVs \citep{Lopes2019,Islam2021} while they use a single simple absorber.

\citet{Orio2009} also uses APEC multi-temperature model to fit the \xmm\ EPIC-pn spectrum. The APEC model consists of 3 temperature components and obtains a spectrum for a collisionally-ionized diffuse gas. For an IP or polar we can expect the presence of a magnetically confined accretion flow onto the magnetic pole (or poles) of the accretor. In this case the impacting material generates a shock heated plasma, which cools towards the surface of the WD, and generates the X-ray emission. This is similar to the multi-temperature APEC model used by \citet{Orio2009}. 

Testing the possibility that the bursts can be explained by micronova requires a magnetic accretor. Therefore, we fit the spectrum using {\sc xspec} version 12.13.0 \citep{Arnaud96}, and specifically use the {\sc cemekl} model \citep{Singh96} which describes the emission from a multi-temperature plasma. The geometry of an accretion stream onto a WD surface can be approximated as cylindrical (see Figure 1 in \citealt{Done1999}). X-rays emitted on the far side of the stream (with respect to the observer) encounter a larger effective column-density to the observer compared to those emitted on the side facing the observer. To account for this we convolve {\sc cemekl} with the absorption model {\sc pwab} \citep{Done1998}, which calculates the emission assuming the non-uniform absorption profile arising from a cylindrical emitter. We further convolve the model with {\sc relfect} \citep{Magdziarz95}, which calculates the reflected emission from neutral material. The use of the {sc reflect} component follows from the presence of the Fe-K$\alpha$ line. Additionally we add two components to model, the fluorescence lines Fe-K$\alpha$ ($\sim 6.4$\,keV), and O\,{\sc vii} ($\sim 0.58$\,keV). The strong O\,{\sc vii} line might be due to the wind from the nova shell, as noted in \citet{Orio2009}. Finally, we also include {\sc tbabs} to account for interstellar absorption as obtained from the visual extinction from Table 1 in \citep{Selvelli2019}, using the optical extinction to H column density relation \citep{Guver2009}. The final {\sc xspec} model is then: {\sc tbabs*pwab*reflect*(cemekl + gaussian(Fe-K$\alpha$) + gaussian(O\,{\sc vii}))}. The best fit model is shown in Figure \ref{fig:SEDx}, and the corresponding model parameters are shown in Table \ref{tab:xspec}.

Table \ref{tab:xspec} shows the resulting best fit parameters. In particular, $T_{\mathrm{max}} = 29.3^{+9.4}_{-7.4}$ keV, significantly lower than $\sim 64$ keV for the multi-temperature APEC model and $\sim 78$ keV for the MKCFLOW model discussed in \citet{Orio2009}. Although $T_{\mathrm{max}}$ is loosely constrained it does not approach the maximum allowed limit of the model. We attribute the main difference between this fit and that of \citet{Orio2009} to the inclusion of the {\sc pwab} component, which represents an approximation for the accretion column. Our new estimate is consistent with the range determined by \citet{Mason2013} using \chandra\ data, which has considerably softer coverage. We note that although  $N_{h, \mathrm{max}}$ is unconstrained its effect on the overall fit is negligible. This particular behaviour of $N_{h, \mathrm{max}}$ is also reported in \citet{Done1998} for BY Cam. Similarly for reflection, which is loosely constrained but with negligible effect on the final value of $T_{\mathrm{max}}$. The possibility of fixing reflection to 1 is also explored. It was found that a fixed value of reflection changes $T_{\mathrm{max}}$ to $T_{\mathrm{max}} = 29.3^{+10.4}_{-7.1}$ keV, however all variations of the temperature are well within the errors, making a fixed value of reflection unnecessary. \citet{Done1998} explored the possibility of fixing $\alpha = 1$, however found that the best fit is obtained for $\alpha < 1$. This is also the case for CP Pup and is indicative of multiple sources of cooling, not only Bremsstrahlung, most likely cyclotron.
%The abundances found in Table \ref{tab:xspec} are higher than those in \citet{Orio2009}. The lower abundances here are expected as they are naturally lowered by the inclusion of gaussian emission lines. However, the O\,{\sc vii} lines are included to improve the fit as the line excess is not included in the model.

To infer the WD mass and mass accretion rate it is here assumed that CP Pup is a magnetic system. This choice was based on partial evidence as presented in \citet{Mason2013} and other works, but also in order to explore the micronova nature of the bursts. From the updated X-ray fit we first isolate the plasma component from the model spectrum, extrapolating it over the energy range for which the model is defined ($5\times 10^{-3}-60$ keV) and integrating it. This results in a bolometric flux of $\sim 1.5 \times 10^{-11}$ erg s$^{-1}$cm${-2}$. Using the \gaia\ inferred distance of $D=780 \pm 11$ pc, this gives $L_{x} \sim 1.1 \times 10^{33}$ erg s$^{-1}$. The WD mass is derived from $T_{\mathrm{max}}$ using Equation 1, relating the shock temperature to the WD mass in magnetic cataclysmic variables, in \citet{Orio2009} following from \citet{Wu2003} and references therein. The resulting mass is $M_{WD} \sim 0.73M_{\odot}$$^{+0.12}_{-0.11}$ with the corresponding radius $R_{WD} \sim 0.011R_{\odot}$$^{+0.001}_{-0.002}$, assuming the mean molecular weight from the abundance in Table \ref{tab:xspec}.

Assuming the inferred $M_{WD}$ is correct, we find a mass accretion rate $\dot{M} = 1 - 2 \times 10^{-10}\,M_{\odot}$\,yr$^{-1}$. The accretion rate is quite low, however estimating the optical luminosity from \gaia\ magnitude yields $\sim 8.8 \times 10^{32}$ erg s$^{-1}$. Therefore estimating the accretion rate from the optical luminosity would lead to similarly low values. As it is likely that the accretion luminosity is not solely constrained to X-rays it is reasonable to assume that the accretion rate above is a lower limit. 

Another method of estimating the accretion rate is through the optical to NIR SED. Using the photometric observations of CP Pup available in vizier\footnote[1]{\label{vizier}\url{https://vizier.cfa.harvard.edu/vizier/sed/}}, the peak of the SED is placed at \xmm\ optical Monitor at $\sim 543$ nm with 2.26 $\pm$ 0.06 mJy. This translates to a luminosity of 2.53 $\pm$ 0.02 $\times$ 10$^{33}$ erg s$^{-1}$ using a distance of 780 $\pm$ 11 pc. Assuming that the luminosity corresponds to the accretion luminosity $L_{acc} = \frac{GM_{WD} \dot{M}}{2R_{in}}$, the estimated accretion rate is $\dot{M} \sim $ 6 $\times$ 10$^{-10}$ - 6 $\times$ 10$^{-9}$ $M_{\odot}$yr$^{-1}$. The range of estimated accretion rate depends on the possible truncation of the inner accretion disc. The values taken into account here estimated $R_{in} \sim R_{WD} - 10 \times R_{WD}$, where $10 \times R_{WD}$ is a value representative of the estimated truncated disc radii in IPs from \citep{Suleimanov2019} for systems of similar mass to CP Pup. Whereas this only represents a very rough estimate of the accretion rate, it corresponds to the same order of magnitude as the X-ray accretion rate. Integrating the entire available SED, with a cut-off after $W_{2}$ \wise\ band, would result in total luminosity of $\sim$ 4 $\times$ 10$^{34}$ erg s$^{-1}$. This would correspond to $\dot{M} \sim $ 9 $\times$ 10$^{-9}$ - 9 $\times$ 10$^{-8}$ $M_{\odot}$yr$^{-1}$, a significantly higher value than previous estimates. The excess may be accounted for if part of the SED is dominated by the donor star as suggested in Section \ref{ss:Porb}. In such a case the approximation of accretion luminosity through the SED peak provides a more accurate estimate. 

It is important to note that we do not have sufficiently high spectral coverage to confidently determine the maximum plasma temperature, $T_{\mathrm{max}}$, which sets the roll-over in the spectrum, and should therefore also affect the derived luminosity. Hence, the above estimates should be treated with caution, noting the model dependence. Furthermore, the accretion rate range above is an lower limit determined by the underlining assumptions.

\begin{table}
    \centering
    \begin{tabular}{c|c|c}
        Parameter & Units & Value \\
        \hline

        \multicolumn{3}{|c|}{----------- {\sc tbabs} -----------} \\
        $N_{H}$ & $10^{22}$\,cm$^{-2}$ & 0.16 \\
        
        \multicolumn{3}{|c|}{----------- {\sc pwab} -----------} \\
        $N_{h, \mathrm{min}}$ & $10^{22}$\,cm$^{-2}$ & $1^{+1625}_{-1} \times 10^{-7}$ \\
        $N_{h, \mathrm{max}}$ & $10^{22}$\,cm$^{-2}$ & $7.34^{+\inf}_{-2.32}$ \\
        $\beta$ & & $-0.86^{+0.04}_{-0.09}$ \\

        \multicolumn{3}{|c|}{----------- {\sc reflect} -----------} \\
        rel\_refl & & $1.06^{+0.94}_{-1.04}$ \\
        Redshift & & 0 \\
        abund & & $0.57^{+0.11}_{-0.11}$ \\
        Fe\_abund & & 1 \\
        $\cos(i)$ & & 0.5 \\

        \multicolumn{3}{|c|}{----------- {\sc cemekl} -----------} \\
        $\alpha$ & & $0.60^{+0.10}_{-0.04}$ \\
        $T_{\mathrm{max}}$ & keV & $29.3^{+9.4}_{-7.4}$ \\
        $n_{h}$ & cm$^{-3}$ & 1 \\
        abund & & $0.57^{+0.11}_{-0.11}$ \\
        Redshift & & 0 \\
        switch & & 1 \\
        norm & $10^{-3}$ & $5.9^{+0.9}_{-0.7}$ \\

        \multicolumn{3}{|c|}{----------- {\sc gaussian} (Fe-K$\alpha$) -----------} \\
        $E$ & keV & 6.4 \\
        $EW$ & keV & 0.07 \\
        $\sigma$ & keV & $10^{-2}$ \\
        norm & $10^{-6}$ & $4.56^{+1.73}_{-1.73}$ \\

        \multicolumn{3}{|c|}{----------- {\sc gaussian} (O\,{\sc vii}) -----------} \\
        $E$ & keV & $0.584^{+0.011}_{-0.010}$ \\
        $EW$ & keV & 0.072 \\
        $\sigma$ & keV & $10^{-3}$ \\
        norm & $10^{-5}$ & $2.2^{+0.4}_{-0.4}$ \\

        \hline

        $\chi^{2}_{\nu}$ & $148.22/159 = 0.93$ 
    \end{tabular}
    \caption{The best fit parameters to the \xmm \ EPIC-pn spectrum of CP Pup, with errors representing $90 \%$ confidence limits. Values with no error were kept frozen during the fitting process, and values with no units are dimensionless. For completeness we also include the switch parameter that were used throughout and determine whether the spectrum is computed via interpolating or pre-computed tables. The abundances used correspond to the 'angr' Table (\url{https://heasarc.gsfc.nasa.gov/xanadu/xspec/manual/node116.html}).}
    \label{tab:xspec}
\end{table}

\section{Discussion and Conclusions}
\label{s:conclusion}

In this section we discuss the interpretation of the nature of the bursts detected in CP Pup. A connection is drawn in Section \ref{ss:micronovae} with micronova, where the details of the model are discussed as well as its limitations and application to CP Pup. The recurrence timescale is determined from the model and compared to that observed from \asassn\ data. As mentioned in section \ref{ss:SEDx}, the interpretation of the nature of the bursts, as well as the accretor mass and accretion rate are here dependent on the assumption that CP Pup is a magnetic system.

%Furthermore, in Section \ref{ss:disc_cluster} possible interpretations of the origin of the persistent signal at $\sim 16.29$ cycles$/$d are discussed. This is then linked to the bursts and the behaviour of the period cluster pre and post burst.

\subsection{The CP Pup bursts as micronova eruptions?}
\label{ss:micronovae}

The energy and shape of bursts shown by CP Pup resemble at least phenomenologically the bursts reported in \citet{Scaringi2022} and \citet{Schaefer2022}. In \citet{Scaringi2022} the bursts are interpreted as micronova bursts, somewhat analogous to Type I X-ray bursts in neutron stars. Micronovae are hypothesised to be the result of a thermonuclear runaway effect in a relatively small magnetically confined accretion column. The objects so far reported to show these bursts are TV Col, EI UMa and ASASSN-19bh, with burst energies ranging from $9 \times 10^{37} $ erg to $1.2 \times 10^{39}$ erg. TV Col in particular has been observed at UV wavelengths during one of the bursts by \cite{SM84}, displaying high ionisation helium and nitrogen lines strengthening during the burst as well as clear P-Cygni profiles suggesting outflow velocities in excess of $3500$km s$^{-1}$. The bursts in TV Col and EI UMa have been compared to those in V1223 Sgr \citep{Hameury2022}. In \citet{Hameury2022} the bursts reported in V1223 Sgr are interpreted by the magnetic-gating accretion instability model as observed in other accreting WDs (\citet{scaringiMVLyr,scaringiTWPic,littlefield22}), but draws no definite conclusion on the nature of bursts in other IPs. Although magnetic gating may explain some observables for V1233 Sgr, it may not explain the large outflows observed by \citet{SM84} in TV Col. The recurrent nova V2487 Oph has also been reported to display fast bursts that appear phenomenologically similar (\citet{Schaefer2022}), but their interpretation has been argued to be attributed to flares due to magnetic recconection events within the accretion disc.

If these are indeed related to the micronova events, then we can attempt to compute the burned/ejected mass during one of these events. The average conversion rate of Hydrogen during the CNO cycle in a standard nova explosion \citep{Bode2008,Jose2020} is $\approx 10^{16}$ erg g$^{-1}$. With this assumption the observed energies convert to ejected masses of $>6.2 \times 10^{-12} M_{\odot}$ and $>3.2 \times 10^{-12} M_{\odot}$ for sectors 7 and 34 respectively, somewhat lower than those reported for TV Col and EI UMa in \citet{Scaringi2022} of $1.8 - 5.8 \times 10^{-11} M_{\odot}$. We note that all these estimates should be considered lower limits as no colour and bolometric corrections have been applied. Hence the energies, and consequently ejected mass would most likely be larger under this interpretation. 

From the micronova model \citep{Scaringi2022a}, the recurrence of the bursts is simply associated to the time it takes for a confined column of material to reach thermonuclear runaway conditions. In the simplest scenario the burst recurrence time is given by $t_{\mathrm{rec}} = \frac{M_{\mathrm{ejected}}}{\dot{M_{\mathrm{acc}}}}$, where $M_{\mathrm{ejected}}$ represents the ejected mass during the burst and $\dot{M_{\mathrm{acc}}}$ the accretion rate of the system. The mass accretion rate estimated from the fit to the \xmm\ spectra is between $1 \times 10^{-10}$ to $2 \times 10^{-10} M_{\odot}yr^{-1}$. This assumes a WD mass of $M_{WD} \sim 0.73M_{\odot}$$^{+0.12}_{-0.11}$, as estimated in Section \ref{ss:SEDx}. Hence the estimated recurrence time would be $t_{\mathrm{rec}} \approx 7 - 23$ days. For comparison the accretion rates quoted above from literature and the one determined in this work are all compared in Table \ref{tab:trecs} alongside with the corresponding recurrence times. Most of the accretion rates in Table \ref{tab:trecs} are also derived from fits to X-ray spectra and in such case the $T_{\mathrm{max}}$ and appropriate model are given as well. The table also notes if the model has a constraint on the $T_{\mathrm{max}}$ parameter and hence the accretion rate. This is judged by the constraint placed on the $T_{\mathrm{max}}$ by the model boundaries of the parameter.

The long-term \asassn\ light curve may also constrain the recurrence timescale of the bursts. Figure \ref{fig:LC} reveals an additional burst not observed by \tess\ after the ones detected in Sector 7 and 34 around $\sim 3005$ BTJD. Along with the 2 bursts observed by \tess\ this suggests a recurrence timescale of $\sim 60$ days. Considering the entire long-term \asassn\ light curve from $\sim 1020$ BTJD to $\sim 3020$ BTJD the recurrence time can be roughly estimated to be $\sim 30 - 60$ days. The lower limit of the range is consistent with the predicted recurrence timescale using the accretion rate from Section \ref{ss:SEDx}. We do note that the data gaps could potentially make this an upper limit on the recurrence time. Beyond the long-term \asassn\ data it is difficult to estimate if the recurrence holds.  

Given the 1942 classical nova eruption, it is plausible to assume that up until then the WD in CP Pup had accreted $\approx 10^{-5}M_{\odot}$ during the preceding centuries, and that this fresh Hydrogen material would be spread across the entire WD surface. On the other hand, from the \asassn\ long-term light curve it appears that fast burst events have been recurring in CP Pup for at least $\sim$4 years. Assuming these events are indeed micronova eruptions, the amount of mass used by the micronova events would then constitute only $\sim$0.005\% of the mass required to trigger the next classical nova. It is of course also possible that micronova events have been occurring since the 1942 nova explosion. In this case, the consumed mass from micronovae would represent about 0.1\% of the required mass to trigger the next classical nova. It is unclear how the recurrence timescales of micronovae and classical nova in the same system relate to each other, but we point out that the two types of explosions do not necessarily need to be mutually exclusive. The magnetic confinement criteria required may not be attained for long enough time to build up enough material to trigger a micronova, allowing lateral spreading of freshly accreted material. The exact conditions for this remain unclear and it is possible that the bursts have not always been present in the system, for example if the conditions for magnetically confined accretion are fulfilled only temporarily. In such a case one or multiple micronova may be triggered. Cessation of micronova events could happen if the magnetic confinement of the accretion column is broken. Whether the accretion column can stay magnetically confined is thought to depend on the combination of multiple parameters, such as the strength of the magnetic field, the column height and footprint area as described in detail in \citet{Scaringi2022a}.

Assuming a micronova interpretation, the lower limit of the recurrence deduced from the long term \asassn\ light curve $t_{\mathrm{rec}} \approx 30$ days, the expected ejected mass would convert to $\sim 3 - 6 \times 10^{-11} M_{\odot}$, consistent with those reported in \citet{Scaringi2022}. This however would imply that the released energy in the \tess\ passband underestimates the bolometric release of energy by a factor of 3 or more. It is however possible that lower mass transfer rates could yield consistent burst recurrence times and energies.

All the recurrence timescales in Table \ref{tab:trecs} are however either shorter or on the lower boundary of the recurrence timescale deduced from the long term \asassn\ light curve. This might be due to the number of necessary conditions for micronova to occur. For example the magnetic pole can change the size of its footprint on the surface of the WD before enough material is accreted for a micronova to be triggered. In such a case the recurrence timescale would become longer. The recurrence timescales could also be underestimated due to the WD mass. The same underestimation can be applicable in the recurrence timescale using accretion rate from other papers, see Table \ref{tab:trecs}. For the mass of the WD to be better constrained however, a wider spectral coverage is necessary, such as that of \textit{NuSTAR}.

\begin{table*}
	\centering
	\caption{Summary of accretion rate estimates for CP Pup from literature and the parameters used to determine it. The recurrent timescale range is computed for the ejected masses above and compared to the results in this work.}
	\label{tab:trecs}
	\begin{tabular}{lcccccccr} % four columns, alignment for each
		\hline
		Paper & Data & Model & $T_{\mathrm{max}}$ & Distance & $M_{WD}$ & $\dot{M}_{\mathrm{acc}}$ & $t_{\mathrm{rec}}$ & Constrained\\
          &  &  & keV & pc & $M_{\odot}$ & $\times 10^{-10}$ $M_{\odot}$yr$^{-1}$ & days &  \\
		\hline
		\citet{Orio2009} & \xmm\ & APEC & 64 & 1600 & >1.1 & $\lesssim 1.6$ & 7.3 $$-$$ 14 & No\\
		\citet{Mason2013} & \chandra\ & VMCFLOW & $36.5^{+19.2}_{-16.3}$ & 1600 & 0.8$^{+0.19}_{-0.23}$ & $4^{+3}_{-1}$ &1.6 $-$ 6.9 & Yes\\
		\citet{Selvelli2019} & $-$ & \citet{Livio1992} & $-$ & 795 $\pm$ 13 & 1.16 $\pm$ 0.2 & $6.0^{+6.6}_{-3.1}$ & 0.9 $-$ 7.8 & $-$\\
        This work & \xmm\ & CEMEKL & $29.3^{+9.4}_{-7.4}$ & 780 $\pm$ 11 & $0.73$$^{+0.12}_{-0.11}$ & 1 $-$ 2 & 7 $-$ 23 & Yes\\
		\hline
	\end{tabular}
\end{table*}

Another important parameter of the micronova model is the fractional area on which material is being accumulated. From the ejected masses $M_{\mathrm{ejected}}$ the fractional area $f$ can be derived from Figure 1 in \citet{Scaringi2022}. For $M_{\mathrm{ejected}}$ and WD mass quoted above the expected $f$ is $\sim 2 - 9 \times 10^{-7}$. This is particularly low for triggering micronova and may point to the bombardment scenario suggested for polars \citep{Frank2002}. In that case, inhomogeneous accretion of dense blobs penetrate into the WD photosphere and radiate most of their energy in a form of soft X-rays. In such a scenario in \citet{Frank2002}, the accretion rate is constrained by $\lesssim 1.6 \times 10^{-11}$ $M_{\odot}$yr$^{-1}$. This is much lower than the estimated accretion rate for CP Pup and argues against the micronova interpretation. The only way out from this potential problem with the model is that the WD mass has been underestimated in the X-ray spectral analysis.

Further photometric, polarimetric and high-time resolution spectroscopic follow-up observations, of the peculiar behaviour displayed by CP Pup will provide additional constraints to unravel its true nature. Specifically the lack of consistent results when it comes to CP Pup, coupled with the lack of a coherent model to explain its behaviour, may warrant a dedicated observing campaign.

\section*{Acknowledgements}
We would like to thank the anonymous referee for the useful and insightful comments that have improved this manuscript. This paper includes data collected with the TESS mission, obtained from the MAST data archive at the Space Telescope Science Institute (STScI). Funding for the TESS mission is provided by the NASA Explorer Program. STScI is operated by the Association of Universities for Research in Astronomy, Inc., under NASA contract NAS 5–26555. This publication makes use of data products from the Wide-field Infrared Survey Explorer, which is a joint project of the University of California, Los Angeles, and the Jet Propulsion Laboratory/California Institute of Technology, funded by the National Aeronautics and Space Administration. MV acknowledges the support of the Science and Technology Facilities Council (STFC) studentship ST/W507428/1. SS is supported by STFC grant ST/T000244/1 and ST/X001075/1. SH acknowledges support from STFC through the studentship ST/V506643/1. KI acknowledges support from Polish National Science Center grant 2021/40/C/ST9/00186. DdM acknowledges financial support from ASI-INAF agreement and INAF Astrofund-2022 grants.

%%%%%%%%%%%%%%%%%%%%%%%%%%%%%%%%%%%%%%%%%%%%%%%%%%
\section*{Data Availability}

The \tess\ data used in the analysis of this work is available on the MAST webpage \url{https://mast.stsci.edu/portal/Mashup/Clients/Mast/Portal.html}. The \asassn\ data \citep{Shappee2014,Kochanek_2017} used for the calibration of \tess\ data is available on the \asassn\ webpage \url{https://asas-sn.osu.edu/}.



\bibliographystyle{mnras}
\bibliography{refs.bib} % if your bibtex file is called example.bib

%\begin{thebibliography}{99}
%\end{thebibliography}

%%%%%%%%%%%%%%%%%%%%%%%%%%%%%%%%%%%%%%%%%%%%%%%%%%

% Don't change these lines
\bsp	% typesetting comment
\label{lastpage}
\end{document}